\documentclass[epj]{svjour}
\usepackage{graphics}
\newcommand{\teff} {T$_{\rm eff}$}
\newcommand{\cd} {d$^{-1}$}
\newcommand{\cy} {y$^{-1}$}
\newcommand{\mjup} {M$_{\rm Jup}$}
\newcommand{\msun} {$_\odot$}

\newcommand{\ms} {m\,s$^{-1}$}

\begin{document}
\title{Observing exoplanets from the planet Earth:
how our revolution around the Sun affects the detection of 1-year periods
\thanks{Contribution to the Focus Point on ``Highlights of Planetary Science in Italy" edited
by P.~Cerroni, E.~Dotto, P.~Paolicchi.}}
\author{Federico Borin\inst{1,2} \and Ennio Poretti\inst{2}
\and Francesco Borsa\inst{2} \and Monica Rainer\inst{2}
}                     
\offprints{E. Poretti}          
\mail{ennio.poretti@brera.inaf.it}
\institute{
Dipartimento di Fisica G. Occhialini, Universit\`a di Milano Bicocca and INFN, 
Piazza della Scienza 3, 20123 Milano, Italy
\and 
INAF-Osservatorio Astronomico di Brera, Via E. Bianchi 46, 23807 Merate, Italy
}
\date{Received: date / Revised version: date}
%
\abstract{
We analysed a selected sample of exoplanets with orbital periods close to 1 year
to study the effects of the spectral window on the data, affected by the 1~\cy\,
aliasing due to the Earth motion around the Sun. 
We pointed out a few cases where a further observational effort would
largely improve the reliability of the orbital solutions.
\PACS{
      {exoplanets: }{planets orbiting other stars}   \and
      {frequency analysis: }{spurious peaks in power spectra}
     } 
} 
\titlerunning{Exoplanets observed from the planet Earth}
\authorrunning{Borin et al.}
\maketitle
\section{Introduction}
\label{intro}

The search for extrasolar planets is not always a straightforward
task. The recent case of $\alpha$~Cen\,B is an example of how stellar
activity and data sampling can conspire in making the detection of a
keplerian signal a controversial exercise. The planet $\alpha$~Cen\,Bb 
would have a mass of only 1.1$\pm$0.1~Earth masses, orbiting a sun-like
star, at only 1.3~pc from the Solar System, with a period of $P_b$=3.24~d only. 
The signal detected in the radial velocity (RV) time series of  $\alpha$~Cen\,B and ascribed
to a keplerian motion has  a semi-amplitude ($K$) of 0.51\,\ms\, (Dumusque et al., 2012).
However, the real nature of the signal was questioned and interpreted
as a ``ghost" signal due to  $\alpha$~Cen\,B activity (Rajpaul, Aigrain, \&
Roberts, 2016).
The signals due to the stellar activity artificially enhanced the peak at 3.24~d present
in the spectral window of the data. Rajpaul et al. emphasized how crucial
is the understanding of every component of a RV time series, including
its spectral window.

There are other subtle examples of artificially induced signals.  The HARPS spectrograph is 
very stable, close to the \ms\, level. However, the RV time series of some stars resulted to
be contaminated by a spurious 1~year signal having an amplitude $K$ of   a few \ms\, (Dumusque
et al., 2015). 
Its artificial nature was imprinted in the phase value, in opposition with the revolution of 
the Earth around the Sun.
It has been explained by the deformation of spectral lines crossing block stitchings of the 
detectors. The spectrum of an observed star is alternatively blueshifted and redshifted due to 
the motion of Earth around the Sun. This annual perturbation can be suppressed {\it a priori}
by either removing the affected spectral lines from the correlation mask  or {\it a posteriori}
by simply fitting a yearly sinusoid to the radial velocity data (Dumusque et al., 2015).

We also remind that the Doppler shifts 
measured from a ground-based observatory 
with respect to an internal  calibration wavelength scale 
have to be transformed into those that would be measured in the barycenter of
the Solar System. This transformation accounts for all the components of
the Earth velocity in the direction of the target due to, e.g., daily rotation,  
Earth-Sun and Earth-Moon systems,~... 
However, we should also consider that high proper-motion stars show 
changing positions and radial velocities with respect to the Solar System
barycenter and these small changes have to be carefully evaluated in the 
era of extremely precise measurements of Doppler shifts (e.g., Wright \& Eastman, 2014).


\section{Sample selection}

\begin{figure*}
\resizebox{0.95\textwidth}{!}{\includegraphics{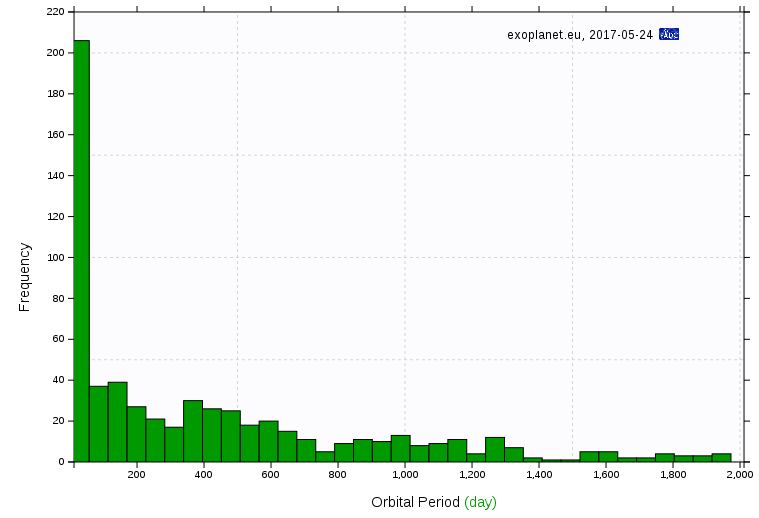}}
\caption{Updated histogram of the orbital periods discovered with the Doppler spectroscopic
technique.
Note the small dip in the distribution around 360~d.} 
\label{isto}       
\end{figure*}
The issues addressed above suggested us a simple project to verify the reliability
of the detections of exoplanets having an orbital period of about 1~year
and discovered with the Doppler spectroscopic technique.
This technique is  much more sensitive
to the aliasing problems than the transit method.
More than 3000 exoplanets were known when we started the project. 
A simple query using {\it The Extrasolar 
Planets Encyclopaedia}\,\footnote{\tt http://exoplanet.eu/} found out 17
candidates with an orbital period in the range from 349.7 to 380.5~d (HD~142b, HD~192699b,
HD~67087b, HD~159868c, HD~210702b, HD~4313b, $\mu$~Leo~b, HD~63765b, HD~17092b,
HD~4732b, HD~96063b, HD~38283b, HD~212771b, HD~73526c, BD+20\,2457b, $\alpha$~Ari\,b,
and HD~20868b). We note that the detection of exoplanets within this range is 
actually more difficult than in others: a clear dip is visible in the frequency
distribution (Fig.~\ref{isto}).

The RV time series could be obtained from the discovery papers or,
in a more useful way, from the NASA Exoplanet Archive\,\footnote{\tt 
https://exoplanetarchive.ipac.caltech.edu/}. Unfortunately, not all the 
discovery's authors put their time series in the paper or in the archive. 
We analysed all the available data and we discuss here the six cases able to provide 
a methodological feedback.

\subsection{HD 142}
\label{hd142}
\begin{figure*}
\centering
\resizebox{0.9\textwidth}{!}{\includegraphics{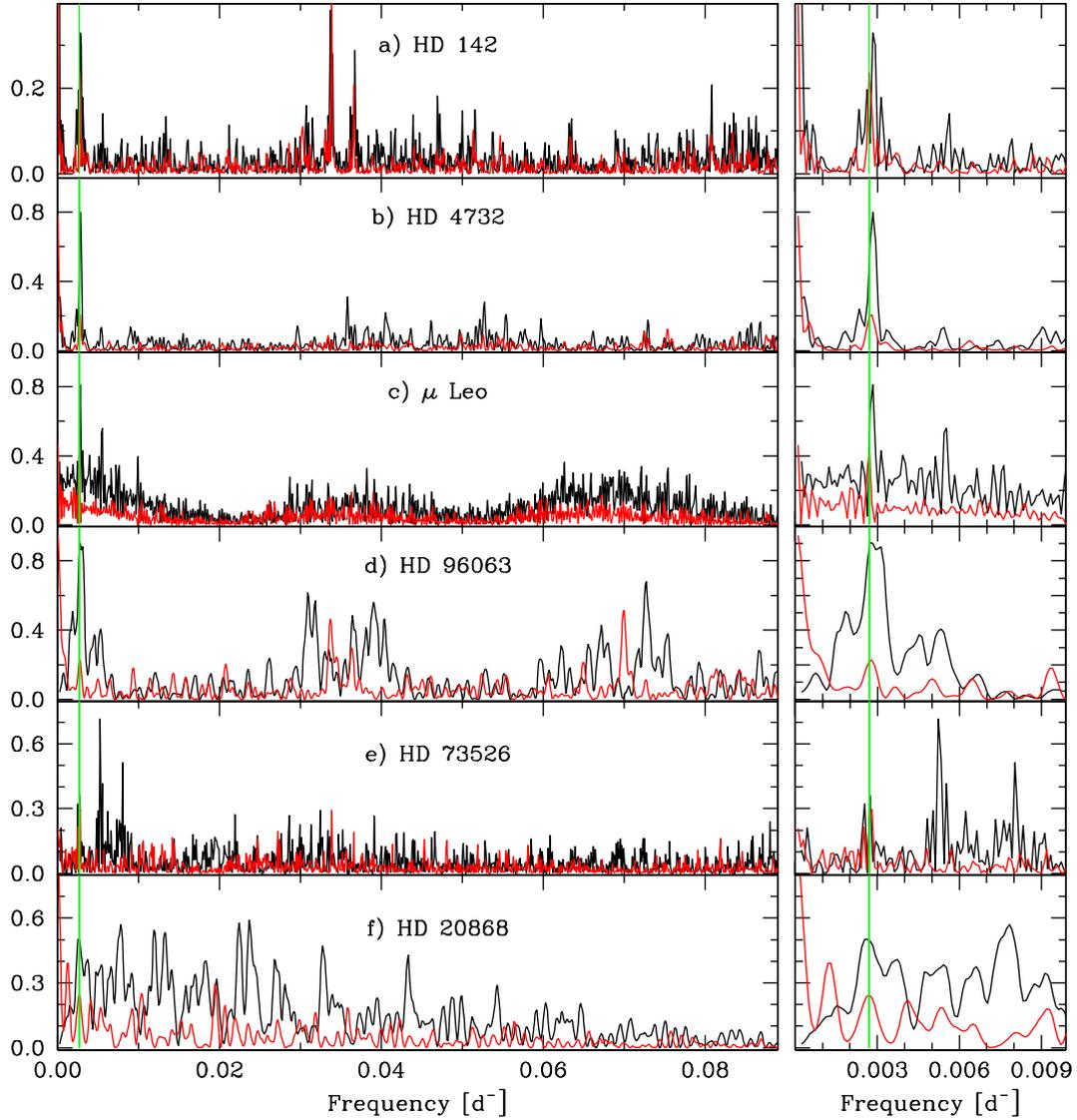}}
\caption{Power spectra ($y$-axis in arbitrary units) of the time series of the six
exoplanetary systems discussed in the paper. In each panel the power spectrum is in black, the spectral window
in red, and a vertical green line shows the position of the 1~\cy\, peak. 
The different widths of the peaks in the power spectra are due to the different
time lengths of the time series.  The panels
on the right side are a zoomed version of those on the left side. They emphasize 
how power spectra and spectral windows are often well superimposed around the 1~\cy\, 
frequency. 
} 
\label{power}       
\end{figure*}

The periods of the two planets orbiting HD~142 (\teff=6025~K, G1IV, 
M=1.15~M\msun) are very intriguing. The planet $b$ ($M_p\sin~i$=1.21~\mjup)
has a period $P_b$=351.1~d, the planet $c$ ($M_p\sin~i$=5.5~\mjup)
a period $P_c$=7900~d (Wittenmyer et al., 2012). 
We performed a careful analysis of the spectral
window, since the authors do not report on it and
the two frequencies 
($f_b$=0.0028~\cd\,  and $f_c$=0.0001~\cd) are separated exactly 
by 1~\cy. 
We were made suspicious by the fact that one period is the 
alias of the other and by the regular structure of the peaks at
low frequencies (Fig.~\ref{power}, panel {\it a}). First harmonics of $f_b$ and
$f_c$ are also detected due to the eccentric orbits. Note that the peak at
0.033~\cd\, is an alias, being reproduced by the spectral window. 

The amplitudes involved here ($K_b$=31.6~\ms, $K_c$=52.6~\ms) 
ruled out the possibility of a misidentification or of an instrumental
effect. Indeed, Wittenmyer et al. 
report a very convincing RV curve over 5000~d and clear phase-folded RV curves 
on the two periods. We just note that the coverage of the $P_b$ orbit 
was not completed, a fundamental step to definitely discard the hypothesis
of a long-term variability due to an activity cycle. 

We also remind the
reader that the planets were discovered around the star HD~142A, but there
is a faint stellar companion (HD~142B, $V$=11.5), with similar high proper motion. 
We noticed that the gravitational bounding of the two stars
has not been taken into account in the evaluation of the long-term variation
of the radial velocity. Moreover, the high proper motion of HD~142A can
also affect the long-term behaviour of the radial velocity.

\subsection{HD 4732}

The presence of a planet around the high proper-motion star HD~4732 
(\teff=4959~K, K0IV, M=1.74~M\msun) was suspected by analysing a RV time series
obtained at the Okayama Astrophysical Observatory.
A continuous increase ranging 100~\ms\, was observed, without a clear
definition of the extrema. Therefore, other RV measurements were
performed at the 3.9\,m Anglo-Australian Telescope and a 
period $P_b$=360~d ($K_b$=47.3$\pm$3.5~\ms) could be inferred  (Sato et al., 2013).
On the basis of five years
of monitoring, Sato et al. also detected a second planet with a period
$P_c$=2732~d ($K_c$=24.4$\pm$2.2~\ms). 

This case is less convincing than the HD~142 one, since the power spectrum
is more noisy (Fig.~\ref{power}, panel $b$). We verified
that none of the two datasets is providing a clear solution when 
analysed separately from the other. When combining them, large parts
of the folded RV curve of planet $b$ remain uncovered (few points on the
steep descending branch).

\subsection{$\mu$ Leo}

A planet around the giant star $\mu$ Leo ($V$=3.88, K2\,III) was detected by
means of 103 RV measurements spanning about 10 years (Lee et al., 2014). 
The resulting period is $P_b$=357.8$\pm$1.2~d. 
The massive planet ($K$=52.0$\pm$5.4~\ms and hence $M_p\sin~i$=2.4~\mjup) is
orbiting at only 1.1~AU from the giant star, whose radii is 16.2~R$_\odot$, i.e., 0.07~AU.
Therefore, the planet is exposed to a large irradiation. 

The spectroscopic  data do not cover the RV curve in a satisfactory way, in particular
at the minimum. A special effort was made in the last observing season to cover
this part of the RV curve, providing a good confirmation of the planetary fit.
The peaks at $f_b$=0.0028~\cd\, and at the first harmonic value
$2f_n$ (eccentric orbit, $e$=0.09$\pm$0.06) stand out above the noise, but 
the power spectrum shows  a general increase of the noise level for $f<$0.01~\cd, due to
the poor spectral window (Fig.~\ref{power}, panel $c$).

It is noteworthy that an independent spectroscopic survey aimed at detected 
pulsation in $\mu$ Leo did not revealed any trace of RV variability (Hekker et al., 2006).
However,
the phase coverage of these RV data was poor on the period very close to 1~year discovered
by Lee et al. and this could explain the non-detection.
The possibility that $\mu$ Leo is a pulsating variable was ruled out by {\sc hipparcos}
photometry (Lee et al., 2014). 
Finally, we note that $\mu$ Leo is another high proper-motion star.

\subsection{HD 96063}

Fourteen RV measurements spanning 1400 days suggested the presence of a planet
orbiting the subgiant star HD~96063 ($V$=8.37, G6) with a period $P_b$=361.1$\pm$9.9~d
and $K_b$=25.9$\pm$3.5~\ms\, (Johnson et al., 2011). 
Such a sampling is prone to annual systematic errors and actually Johnson et al.
carefully verified if the barycentric correction was a possible error source. No correlation
was found and no similar signal was detected in other targets. The few measurements
per year make the power spectrum a bit noisy, with large structures (Fig.~\ref{power}, panel $d$).
The phase coverage is also very poor.

\subsection {HD 73526}

The real existence of the 2:1 resonant exoplanetary system ($P_b=$187.5~d, $P_c$=376.9~d) 
around HD~73526 ($V$=4.1, G6V) has not been easy to ascertain (Tinney et al., 2006), 
but there now is a good 
theoretical background supporting it (Wittenmyer et al., 2014). 
The fact that the second period is close
to 1~year did not help in securing a good phase coverage. 
The power spectrum clearly detects the highest peak at $f_b$=0.0054~\cd, while the peak at 1~\cy\,
is much lower (Fig.~\ref{power}, panel $e$).

Both spectroscopic orbits have
large amplitudes ($K_b$=83.0~\ms\, and $K_c$=62~\ms). We notice that the orbital solution 
could be improved by planning new measurements covering the extrema: in few
occasions only the highest maxima and the lowest minima have been observed, and almost
all times with few measurements (Fig.~1 in Wittenmyer et al., 2014). This weakens  the reliability
of the periods, as demonstrated by the changes in the values of the parameters in the
solutions proposed first by Tinney et al. (2006) 
and then by Wittenmyer et al. ( 2014). 

\subsection{HD 20868}

The detection of the planet orbiting  HD~20868 (Moutou et al., 2009) 
is clearly unquestionable.
Despite a period close to 1~year ($P_b$=380.85~d, the longest in our sample) and a very eccentric orbit ($e$=0.75),
the HARPS measurements  covers the RV  abrupt changes around the periastron in a very satisfactory
way.  The maximum RV values and the amplitude ($K_b$=100.34$\pm$0.42~\ms) are also well constrained by 
the dense monitoring secured during the last passage at the periastron. 
The resulting power spectrum (Fig.~\ref{power}, panel $f$)
shows the complex structure due to the several harmonics (not observed in the spectral
window) we have to consider to fit the non-sinusoidal RV curve.

\section{Conclusions}

Our analysis of the exoplanets with orbital periods of about 1~y did not revealed any spurious
results, but just a few cases (HD~4732b, HD 4732c, $\mu$~Leo~b, HD~96063b) deserving more
observations. All the stars are quite bright and the requested time sampling should try to
cover the largest part of the year as possible, with a cadence of 2-3 measurements per month.
Such a task can be accomplished by long-term programs, like the {\it Global Architecture of Planetary
Systems} (GAPS) with HARPS-N at the Telescopio Nazionale Galileo (Poretti et al., 2016). 
However, the scheduling should take care to limit the months without data acquisitions to those with 
the targets in close conjunction with the Sun. This is the gold rule to minimize the 1~\cy\, alias and
to ensure an almost perfect satisfactory coverage of the folded RV curves.
\begin{acknowledgement}
The whole team of authors acknowledges support from {\it Istituto Nazionale di Astrofisica} (INAF) through 
the {\it Progetti Premiali} funding
scheme of the Italian Ministry of Education, University, and Research.
\end{acknowledgement}
%
%
%
%

%

\begin{thebibliography}{}
%
%
\bibitem{ccd} Dumusque, X., Pepe, F., Lovis, C., \&  Latham, D.W., ApJ, \textbf{808}, (2015), 171

\bibitem{dumu} Dumusque, X., Pepe, F., Lovis, C., et al.,  Nature, \textbf{491}, (2012) 207


\bibitem{johnson} Johnson, J.A., Clanton, C., Howard, A.W., et al., ApJ Suppl. Ser., \textbf{197}, (2011), 26

\bibitem{hekker} Hekker, S., Reffert, S., Quirrenbach, A., et al., A\&A, \textbf{454}, (2006), 943

\bibitem{lee} Lee, B.-C., Han, I., Park, M.-G., et al., A\&A, \textbf{566}, (2014), A67

\bibitem{moutou} Moutou, C., Mayor, M., Lo Curto, G., et al., A\&A, \textbf{496}, (2009), 513

\bibitem{gaps} Poretti, E., Boccato, C., Claudi, R., et al., MemSAIt, \textbf{87}, (2016), 141

\bibitem{ghost} Rajpaul, V., Aigrain, S., Roberts, S., MNRAS, \textbf{456}, (2016), L6

\bibitem{sato} Sato, B., Omiya, M., Wittenmyer, R.A., et al., ApJ, \textbf{762}, (2013), 9

\bibitem{tinney} Tinney, C.G., Butler, R.P., Marcy, G.W., et al., ApJ, \textbf{647}, (2006), 594

\bibitem{witte} Wittenmyer, R.A., Horner, J., Tuomi, M., et al., ApJ, \textbf{753}, (2012), 169

\bibitem{reso} Wittenmyer, R.A., Tan X., Hoi Lee, M., et al., ApJ, \textbf{780}, (2014), 780

\bibitem{wright} Wright, J.T. \&   Eastman, J.D., PASP, \textbf{126}, (2014), 838
\end{thebibliography}
%

\end{document}